# Pashen-Back effect for Plasma Diagnostics

**Koryun Oganesyan,**[1,2,*] **, Krzysztof Dzierzega,**[3] **, Ashot Gevorgyan,**[4] **and Peter Kopcansky**[2]

[1] *Alikhanyan National Lab, Yerevan Physics Institute, Alikhanyan Br.2, 0036, Yerevan, Armenia*
[2] *Institute of Experimental Physics SAS, Watsonova 47, 04001 Kosice, Slovakia*
[3] *Marian Smoluchowski Institute of Physics, Jagiellonian University, ul. Łojasiewicza 11, 30-348 Krak´ow, Poland*
[4] *Institute of High Technologies and Advanced Materials, Far Eastern Federal University, 10 Ajax Bay, Russky Island, 690922    Vladivostok, Russia*
*\*bsk@yerphi.am*

**Abstract:** The possibility of determining the magnitude of neutral atom density in hydrogen plasma was investigated based on the Pashen – Back effect and resonant Faraday rotation of the polarization plane of light by residual neutral atoms in the plasma (or by specially introduced impurity atoms ). In strong magnetic fields when the Zeeman shift $\mu_0 H$ is greater than the distance of the hyperfine structure energy levels the Paschen-Bak effect is observed. In this case, it is no longer possible to speak about the independence of the splitting of each level of a given multiplet term.
This method combines the Pashen-Back effect with the effect of resonant Faraday rotation of polarization plane of the probe signal.

## 1. Introduction

Magneto-optical (MO) effects, magnetically induced changes in light intensity or polarization upon reflection from or transmission through a magnetic sample, were discovered over a century and a half ago [1–3]. Initially, they played a crucial role in unveiling the fundamentals of electromagnetism and quantum mechanics. However the broader relevance and wide-spread use of MO methods, remained quite limited until the 1960s owing to a lack of suitable, reliable and easy-to-operate light sources. The advent of laser technology and the availability of other novel light sources have led to an enormous expansion of MO measurement techniques and applications that continues to this day. Magneto-optics (MO) has wide applications in spectroscopy, chemistry, plasma diagnostics, magnetic field sensors, theranostics, pharmaceutic, and magnetometry [4].

Magneto-optics refers to changes in the properties of light when it is transmitted or reflected in the presence of a magnetic field (externally applied or from a magnetized material medium) [5]. Magneto-optics is widely used in plasma polarimetric diagnostics [ 6−13 ].

When plane polarized light passes through a plasma located in a magnetic field, the polarization plane of the light is rotated. When the propagation direction of the light coincides with the direction of the magnetic field, the rotation of the polarization plane is a result of the Faraday effect. If the frequency of the light is close to the frequency of an atomic (or ionic) transition, then the change in the polarization of the light is caused by a resonant interaction of the light with the atoms (or ions).

An effect on spectral lines obtained when the light source is placed in a very strong magnetic field, first explained by F. Paschen and E. Back in 1921 [14]. Hyperfine Pashen-Back regime was investigated experimentally in Rb atomic medium in [15], high resolution magnetic field measurements in hydrogen and helium plasmas using active laser spectroscopy were reported in [16].

One of the main tasks set before plasma researchers is the problem of using thermonuclear energy. At the first stages of development, the main attention was paid to the determination of such plasma parameters as the concentration of charged particles and (electrons and ions ), their temperatures and , the degree of ionization, etc. Now more and more attention of researchers is attracted by the problem of determining the residual density of the neutral component of the plasma and the magnitude of the magnetic field.

In the present Letter we investigate theoretically the Faraday effect in an hydrogen plasma in the hyperfine Paschen–Back regime for plasma diagnostics when the Zeeman interaction is larger than the hyperfine splitting.

## 2. Splitting of energy levels

First we examine the case in which light interacts with neutral atoms in a hydrogen plasma. Let us suppose that the frequency of the light is close to that of the $n=2 \to n=3$ transition $H_\alpha$ ($\lambda = 656,3 nm$) of atomic hydrogen ($n$ is the principal quantum number). We suggest to use ruby laser with operation wavelength $\lambda = 694,3 nm$. The Faraday effect is closely coupled to the splitting and shifting of the energy levels of hydrogen atoms in a magnetic field (the Zeeman and Pashen-Back effects). Typically, hydrogen plasma is in a magnetic field $H \geq 10^3 Gs$. For such fields the magnetic splitting parameter $\mu_0 H$ in hydrogen becomes comparable to the fine structure interval ($\delta E \approx 10^{-1} cm^{-1}$) so that the Pashen-Back effect appears.

Taking the $z$ axis to be along the magnetic field, we now find the energy splitting of the term with spin $S=1/2$ for Pashen-Back effect [17]. The new value of the energy in a strong magnetic field ($\mu_0 H \geq \delta E$) are

$$E_{\pm(L+1/2)} = \delta E \pm \mu_0 H(L+1), \quad M = \pm(L+1/2)$$

$$E_{\pm M_I} = \mu_0 H(M_I \pm 1/2) + \frac{\delta E}{2} \pm \frac{M_I \delta E}{2L+1}, \quad (1)$$

$$M = L - 1/2, ..., -(L-1/2)$$

where $M_I = (L-1/2),...,(L+1/2)$; $L$ is the orbital angular momentum; $\delta E$ is the separation between the components of doublet in the absence of a field.

We assume that the signal propagates along the magnetic field direction ($z$ axis) and is polarized in the $xy$ plane. It is convenient to represent the polarization of this signal in the form of the sum of two circularly polarized components $E_\pm = E_x \pm iE_y$. In accordance with the selection rules $\Delta L = \pm 1$, $\Delta M_L = \pm 1$ and $\Delta M_S = 0$, we found that for each of the circular components of the polarization there are ten transitions (Table.1 Supplemental Material).

Note $\Delta \omega_0 = \omega - \omega_0$ is a resonance frequency difference between the frequency of the light $\omega$ and the frequency of the $2S_{1/2} \to 3P_{1/2}$ transition, $\delta E_i$ (i = 1, 2, 3) are the finite structure intervals given by (see Table.1)

$$\delta E_1 = E_{2P_{3/2}} - E_{2P_{1/2}} = 0.366\,cm^{-1}$$

$$\delta E_2 = E_{3P_{3/2}} - E_{3P_{1/2}} = 0.108\,cm^{-1}$$

$$\delta E_3 = E_{3D_{5/2}} - E_{3D_{3/2}} = 0.036\,cm^{-1}.$$

Table 1. Splitting of energy levels in strong magnetic fields when $\mu_0 H \gg \delta E$ (the Pashen-Back regime)

| n | l | j | | finite structure intervals |
|---|---|---|---|---|
|   |   |   |   |   |
| 3 | 2 | 5/2<br>3/2 | $3D_{5/2}$<br>$3D_{3/2}$ | $\delta E_3 = 0.036\,cm^{-1}$ |
|   | 1 | 3/2<br>1/2 | $3P_{3/2}$<br>$3P_{1/2}$ | $\delta E_2 = 0.108\,cm^{-1}$ |
|   | 0 | 1/2 | $3S_{1/2}$ |   |
| 2 | 1 | 3/2<br>1/2 | $2P_{3/2}$<br>$2P_{1/2}$ | $\delta E_1 = 0.366\,cm^{-1}$ |
|   | 0 | 1/2 | $2S_{1/2}$ |   |
| 1 | 0 | 1/2 | $1S_{1/2}$ |   |

## 3. Polarization change

The change in polarization of a probing light signal that propagates through this resonant medium is examined.
The reduced propagation equations for slowly varying amplitudes of $E_+$ and $E_-$ are found assuming that the atoms have an isotropic Maxwellian velocity distribution [7],

$$\frac{dE_\pm}{dz} = -\frac{\chi}{2}(A_\pm + iB_\pm)E_\pm, \quad (2)$$

where $\chi = \frac{6\pi^{3/2}e^2 \sqrt{\ln 2} \Delta N}{mc\gamma_{Dop}}$ ; $\Delta N = \frac{N_2}{8} - \frac{N_3}{18}$ is the difference between the population of a single magnetic sublevel in the lower state $n = 2$ and that of one in the upper state $n = 3$ ; $A_\pm = \sum_{k=1}^{10} r_k^\pm f_k^\pm u(\alpha_k^\pm, \beta)$ and $B_\pm = \sum_{k=1}^{10} r_k^\pm f_k^\pm v(\alpha_k^\pm, \beta)$, where the coefficients $r_k^\pm$ are Lande coefficients and $f_k^\pm$ are the oscillator strengths. The probability integrals [18]

$$u(\alpha, \beta) = \frac{1}{\pi} \int_{-\infty}^{+\infty} \frac{\beta e^{-t^2} dt}{(\alpha - \beta)^2 + \beta^2},$$

$$v(\alpha, \beta) = \frac{1}{\pi} \int_{-\infty}^{+\infty} \frac{(\alpha - t)e^{-t^2} dt}{(\alpha - \beta)^2 + \beta^2} \quad (3)$$

are taken at the points $\alpha_k^\pm = \frac{2\sqrt{\ln 2}\,\varepsilon_k^\pm}{\gamma_{Dop}}$ and $\beta = \frac{\gamma\sqrt{\ln 2}}{\gamma_{Dop}}$; $\gamma$ is the homogeneous width of the spectral line; $\gamma_{Dop} = \frac{2\omega}{c}\sqrt{(\ln 2)\frac{2kT}{m}}$ is the inhomogeneous Doppler width; and the resonant frequency difference $r_k^\pm$ are given.

The complete pattern of splitting and shifts of the energy levels of hydrogen in a magnetic field are obtain applaying the results of Eq.(1) to the $S_{1/2}$ and $P_{1/2}, _{3/2}$ terms of the $n = 2$ level and to the $S_{1/2}$, $P_{1/2}, _{3/2}$, and $D_{3/2}, _{5/2}$ terms of the level $n = 3$. It should be noted that of the eight magnetic sublevels of the $n = 2$ level, only two sublevels remain degenerate in a magnetic field; of the 18 magnetic sublevels of the $n = 3$ level, only four sublevels remain degenerate in pairs. Each level with a given $n$ must be split into $n$ sublevels corresponding to the values of $j$ from $1/2$ to $n - 1/2$. The energies of each pair ($2S_{1/2}, 2P_{1/2}$), ($3S_{1/2}, 3P_{1/2}$), ($3P_{3/2}, 3D_{3/2}$), ... of levels with the same values coincide according to Dirac's theory.

The shifts in the levels typically lie within the Doppler width; thus, for $T \approx 100\,eV$ and $H \approx 10^4\,Gs$, the parameter $\mu_0 H \approx 1\,cm^{-1}$ and, the Doppler width $\gamma_{Dop} \approx 10\,cm^{-1}$ (for $T \approx 1\,eV$, the Doppler width $\gamma_{Dop} \approx 1\,cm^{-1}$). The hyperfine Pashen-Back Faraday effect occurs when the Zeeman shift $\mu_0 H$ is of the order of or greater than the Doppler width $\gamma_{Dop}$. Selective reflection spectroscopy can be used to eliminate Doppler broadening in some cases [20].

From equations (2) and (3) it is clear, that a circular dichroism ($A_+$ and $A_-$) and different circular dispersion of the refractive index ($B_+$ and $B_-$) appear in the medium because of the splitting and shifting in the energy levels of the atoms in a magnetic field,.

It is possible to follow the change in the polarization of the probe signal according to Eq.(2). If the wave is polarized along the x-axis at the input point, the y-component of the polarization will also exist at the exit point. Power $P_y(z)$ is related to the input power $P_x(0)$ by

$$P_y(z) = \frac{P_x(0)}{4} \frac{\chi^2 z^2}{4} \times \left[ |A_- - A_+|^2 + |B_- - B_+|^2 \right] e^{-\chi A_+ z}. \quad (4)$$

The amount of the polarization plane rotation for typical experimental conditions ($T; 1eV$, $H; 30\,kGs$, $\Delta\omega_0 = 0$, $\chi; 10^{-14}\Delta N$) is estimated. In equilibrium the difference in populations can be found with the formula $\chi; \Delta N = 0.5 \times 10^{-17} N_e N_i$, where $N_e$ is the electron density and $N_i$ is the density of hydrogen atoms in the ground state $n=1$ [20].

Substituting the numerical values of the probability integrals [19], we find the angle of rotation of the polarization plane at the exit side of a $z = 10\,cm$ thick medium to be

$$\varphi = \sqrt{\frac{P_y(z)}{P_x(0)}} = 17.5 \times 10^{-32} N_e N_i. \quad (5)$$

In modern experiments, it is possible to measure the angle of rotation of polarization plane with an accuracy of up to $\varphi \approx 10^{-5}\,rad$. Thus it follows from (5) that the minimum density of hydrogen atoms required for measuring a magnetic field $H = 30\,kGs$ is $(N_i)_{min} = 6 \times 10^{11} - 6 \times 10^{12}\,cm^{-3}$ (for $N_e = 10^{13} - 10^{14}\,cm^{-3}$). Our calculations show that the angle of rotation of the polarization plane increases with the magnetic field Fig 1. (a), but that the relationship between $\varphi$ and $H$ is not strictly linear Fig.1 (a), (b). This is a result of the averaging of the complex refractive index over Doppler line profile (see Fig.1).

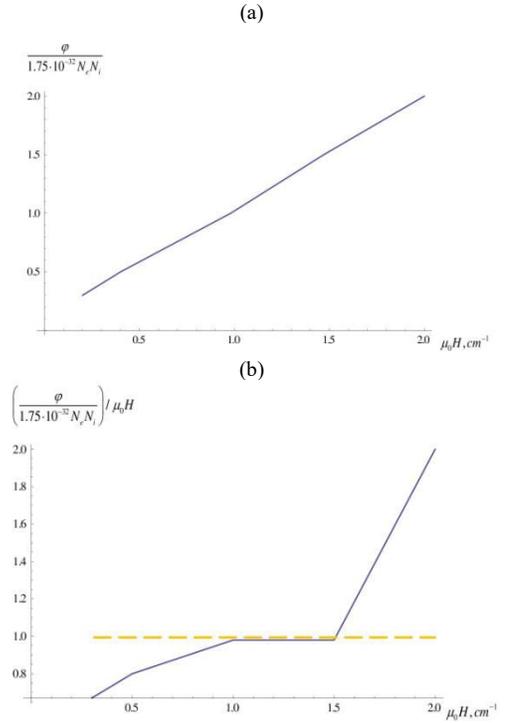

Fig.1. (a) The dependence of polarization plane rotation angle on magnetic field strength; (b) the dependence of polarization plane rotation angle and hyperfine splitting parameter $\mu_0 H$ ratio on magnetic field strength. The angle of rotation of the polarization plane increases with the magnetic field Fig 1. (a), but the relationship between $\varphi$ and $H$ is not strictly linear. It is illustrated in Fig. 1 (b). Orange dashed line corresponds to strict linear case.

### 4. The role of Stark broadening

In plasma, one of the main contribution to the impact broadening is the interaction with electrons and ions. If an atom has a non-zero average dipole moment, the line broadening occurs as a result of the linear Stark effect. This situation is realized for hydrogen. which, due to random degeneracy in the orbital number $l$ has a nonzero dipole moment. For other atoms, a quadratic Stark effect occurs : a passing particle polarizes the atom and then interacts with it. Consider the linear Stark effect. Note that electrons and ions lead to impact broadening under interesting us conditions. The applicability criterion is the value of the parameter $h = N\left(\pi \frac{C_2}{v}\right)^3 \ll 1$ [9], where $C_2 = \frac{3}{2} z n(n-1)\frac{h}{m}$ is the Stark constant ($z$ ion charge, $n$ main (principal) quantum number of the broadened hydrogen level), $N$ - density of neutral atoms, $m$ electron mass. In the interested us conditions, the speed of ions $v_i \sim 10^5\,cm/s$, electron velocities $v_e \sim 10^7\,cm/s$, density of ions and electrons $N_e : N_i : 10^{13} \div 10^{14}\,at/cm^3$, $C_2 \sim 1$. Parameter $h$ for ions $h_i = 10^{-2} \div 10^{-1}$ for electrons $h_e = 10^{-8} \div 10^{-7}$. Thus, as for ions, and electrons, impact broadening is realized for $h_{i,e} \ll 1$.

The most significant Stark splitting is for hydrogen, which has a linear Stark effect. Let us estimate the value of the Stark shift in hydrogen using the formula [9]

$$\gamma_{St} = \frac{3}{2} n(n-1) e a_0 F_0 \qquad (6)$$

where $F_0 = 2.6 e N_i^{2/3}$ - normal field of Holzmark, $a_0$ the Bohr radius. At $N_i : 10^{13} \div 10^{14} at/cm^3$, the value is $\gamma_{St} \sim 10^2 \div 10^{-1} cm^{-1}$.

The magnitude of the Stark shift is quite large $\gamma_{st} \sim 10^2 \div 10^{-1} cm^{-1}$, it approaches the Zeeman broadening $\mu_0 H \approx 1 cm^{-1}$. However, it is important to emphasize, that the linear Stark effect does not remove the degeneration of $M$, and therefore will not affect the change in the polarization of the radiation.

## 5. Conclusion

This method can also be used to determine the density of neutral component in a plasma when the magnetic field is known.
In our earlier studies [21,22] the average and local magnetic field and neutral atom density were found in plasma.
It should be noted that this consideration gives an averaged picture of density of hydrogen atoms similar to that in [21]. The local diagnostics should be done using stimulated rotation of polarization plane of a weak signal in the field of intense one in the presence of external magnetic field similar to Zeeman effect considered in [22].
It should be emphazied that this method is applicable for plasma with temperature from $T \approx 1 eV$ up to tokamak temperatures $T \approx 100 eV$, atom minimal density $N_i = 6 \times 10^{11} - 6 \times 10^{12} cm^{-3}$, electron density $N_e = 10^{13} - 10^{14} cm^{-3}$.

**Acknowledgments.** KBO thanks EU NextGenerationEU through the Recovery and Resilience Plan for Slovakia under project No. 09I03-03-V01-00052, by the Slovak Academy of Sciences, in the framework of project VEGA 2/0061/24 and Development Agency project No. APVV-22-0060 MAMOTEX. AHG thanks the Foundation for Theoretical Physics and Mathematics "BASIS (Grant No. 21-1-1-6-1).